\documentclass[12]{article}
\usepackage[utf8]{inputenc}

\usepackage{graphicx}
\usepackage{float}

\newcommand\pubnumber{}
\newcommand\pubdate{\today}

\def\napoli{
Cukurova University, Adana, TURKEY}
\def\support{\footnote{Work supported by COST Action EURONuNeT (CA15139)}}

\def\Title#1{\begin{center} {\Large #1 } \end{center}}
\def\Author#1{\begin{center}{ \sc #1} \end{center}}
\def\Address#1{\begin{center}{ \it #1} \end{center}}

\newcommand\pubblock{\rightline{\begin{tabular}{l} \pubnumber\\
         \pubdate  \end{tabular}}}
\newenvironment{Abstract}{\begin{quotation}  }{\end{quotation}}
\newenvironment{Presented}{\begin{quotation} \begin{center} 
             PRESENTED AT\end{center}\bigskip 
      \begin{center}\begin{large}}{\end{large}\end{center} \end{quotation}}

\begin{document}
\begin{titlepage}
\pubblock

\vfill
\Title{Combining Forces for a Novel European Facility
for Neutrino-Antineutrino Symmetry-Violation
Discovery (EuroNuNet)}
\vfill
\Author{ Gul Gokbulut
\support}
\Address{\napoli}
\vfill
\begin{Abstract}
I describe a Cost Action (CA CA15139) and a Horizon2020 (H2020) Project (ESSnuSB) to study the feasibility of using the European Spallation Source (ESS) to produce the world’s most intense neutrino beam and measure the parameters of neutrino oscillation, leading to the determination of the value of $\delta$\textsubscript{CP}.
\end{Abstract}
\vfill
\begin{Presented}
NuPhys2017, Prospects in Neutrino Physics
Barbican Centre, London, UK,  December 20--22, 2017
\end{Presented}
\vfill
\end{titlepage}
\def\thefootnote{\fnsymbol{footnote}}
\setcounter{footnote}{0}
\section{Introduction}
The Big Bang should have created equal amounts of matter and antimatter; however today matter remains in the Universe. The European Spallation Source neutrino Super Beam’s (ESSnuSB) main
objective is to demonstrate the feasibility of using the European Spallation Source (ESS) \cite{The European Spallation Source} proton linac in Lund to produce the world's most intense neutrino beam simultaneously with the 5 MW proton
generation for neutron production and measure the parameters of neutrino oscillation, leading to the determination of the value of $\delta$\textsubscript{CP}. Once it is constructed ESSnuSB aims to explain the matter/antimatter
asymmetry in the Universe. 
The Cost Action EuroNuNet (Combining forces for a novel European facility for neutrino-antineutrino symmetry-violation discovery ) is formed to bring together the European
neutrino physicists interested in this concept and to impact the priority list of High Energy Policy makers and of funding agencies to this new approach to the experimental discovery of leptonic CP violation. A Horizon 2020 Project (ESSnuSB) to study the feasibility of this approach has been approved by European Commission and started from January 2018.

\section{The ESS Proton Linac}
\begin{figure}[htb]
\includegraphics[width=\linewidth]{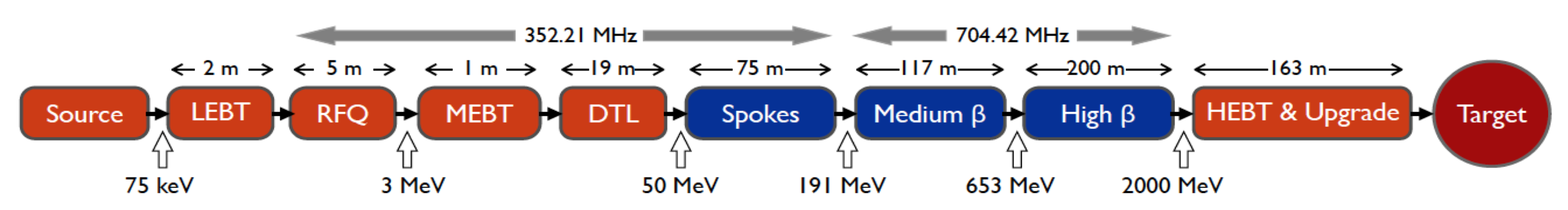}
\caption{ESS Proton Linac}
\label{fig:1}
\end{figure}

Figure \ref{fig:1} shows the ESS Linac which will be a copious source of spallation neutrons with 2 GeV proton beam, 5 MW average beam power, 125 MW peak power, 14 Hz repetition rate and 4 $\%$ duty cyle.
During the future upgrades, an empty space at the end of the linac could allow to increase the proton energy up to 3.6 GeV.
It is known that the second oscillation maximum provides more sensitivity to CP violation. In order to reach the second oscillation maximum, either the experiment baseline should be increased or neutrinos with lower energy should be used or both as regard to the experiments operated at the first oscillation maximum \cite{Marcos Dracos}. By increasing the baseline a decrease of the statistics is expected caused by the solid angle and if the neutrino energy goes below $~$1 GeV, rapidly decreasing neutrino interaction cross-sections are expected. This is why very intense neutrino beams are needed to use the second oscillation capabilities.  European Spallation Source (ESS) \cite{E. Baussan}  could provide such neutrino beams.

\section{Neutrino Species}
\begin{figure}[htb]
\includegraphics[width=\linewidth]{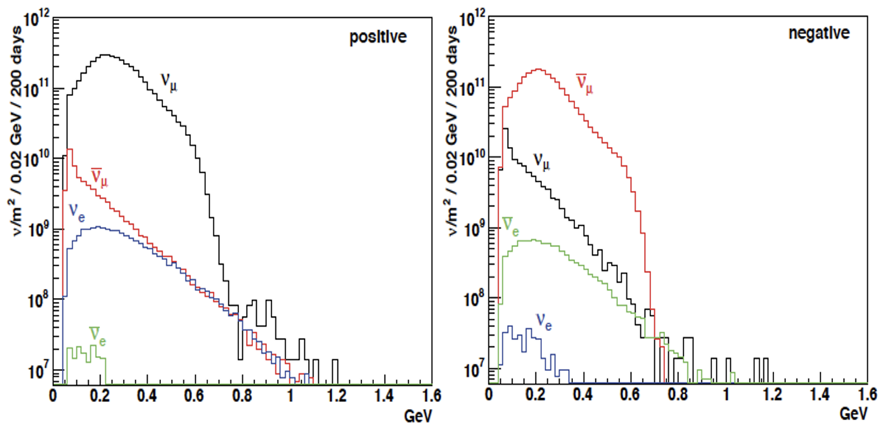}
\caption{Neutrino energy distribution at a distance of 100 km on–axis from the target station, for 2.0 GeV protons and positive (left,
neutrinos) and negative (right, anti–neutrinos) horn current polarities, respectively.}
\label{fig:2}
\end{figure}

Using the ESS proton beam and the target station, the neutrino and anti–neutrino energy
distributions of Figure \ref{fig:2} can be obtained at a distance of 100 km from the target using 2 GeV protons. Neutrinos or anti–neutrinos are obtained by focussing $\pi^{+}$  or $\pi^{-}$ respectively according to the direction of the electric current in the horn. Current studies based on simulations show almost pure $\nu$\textsubscript{$\mu$} with small $\nu$\textsubscript{e} contamination can be established.

\section{Underground Detector and It's Location}
A megaton Water Cherenkov far detector with fiducial volume of order of 500 kt that is similar to MEMPHYS will be used \cite{Marcos Dracos 2}.  
 Figure 3 shows several candidate mines that could host this detector. Garpenberg (located at 540 km) and Zinkgruvan (located at 360 km) which are in Sweden are considered as the best candidates.
The fraction of the full $\delta$\textsubscript{CP} range within which CP violation can be discovered at 5 $\sigma$ (3 $\sigma$ ) significance is above 40$\%$(67$\%$) in the range of baselines from 300 km to 550 km and has the maximum value of 50$\%$ (74$\%$) at around 500 km for 3.0 GeV.
 
\begin{figure}[H]
\centering
\includegraphics[height=3in]{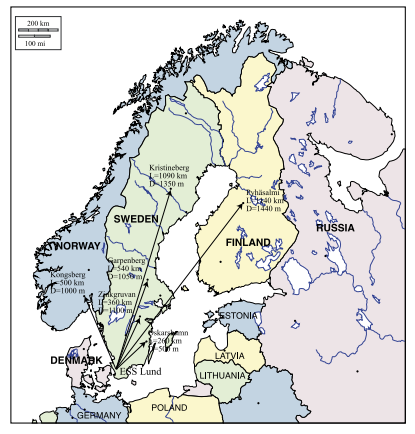}
\caption{Location of the ESS site and the sites of several deep ($>$ 1000 m) underground mines. The distance (L) from
ESS-Lund and the depth (D) of each mine is indicated below the name of the mine.}
\label{fig:3} 
\end{figure}
\begin{figure}[H]
\centering
\includegraphics[height=2.5in]{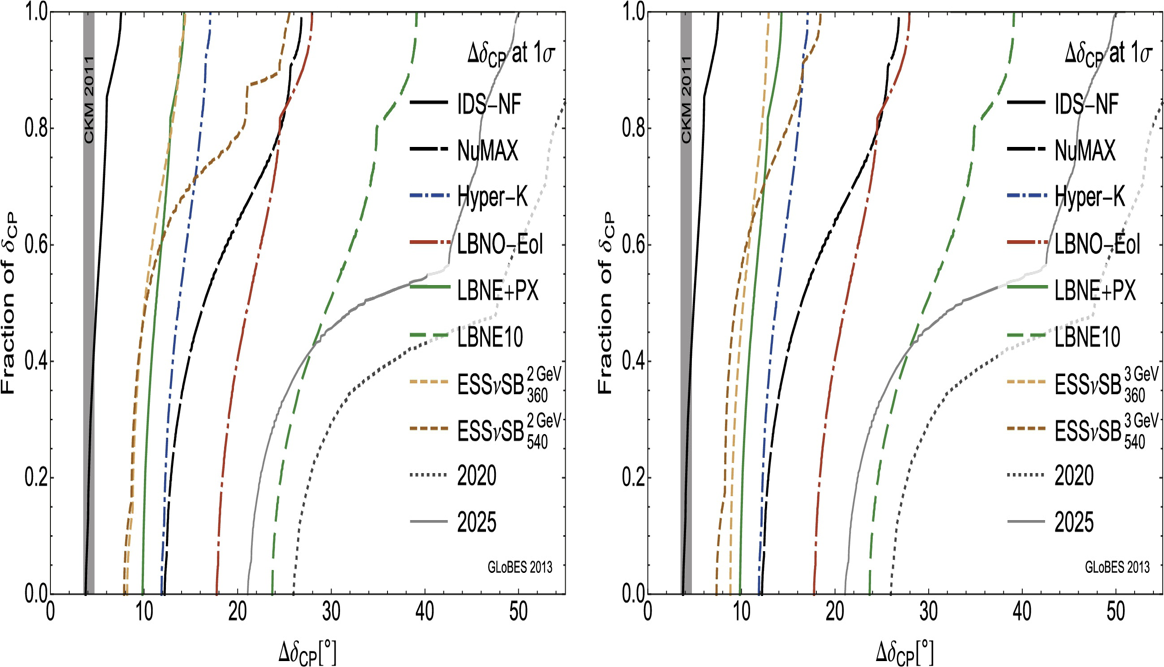}
\caption{Fraction of the full $\delta$\textsubscript{CP} range for which an error $\Delta$$\delta$\textsubscript{CP} (at 1 $\sigma$  , 1 d.o.f. ) or better could be achieved. For ESSnuSB
the two baselines of 360 km and 540 km and two proton energies (2.0 GeV on left and 3.0 GeV on right) have been used.}
\label{fig:4} 
\end{figure}

\section{Comparison with other Neutrino Experiments}
After discovering leptonic \textsubscript{CP} violation (i.e $\delta$\textsubscript{CP} $=$ 0$^{\circ}$, 180$^{\circ}$) the next goal will be to measure $\delta$\textsubscript{CP} as precisely as possible. The precision that can be reached in this measurement represents an additional discriminating criterion among the experimental options. Fig. 4 \cite{E. Baussan} presents a comparison with other projects of the fraction of the full $\delta$\textsubscript{CP} range covered for which an error $\Delta$$\delta$\textsubscript{CP} or better could be achieved. Here also it can be seen that only the Neutrino Factory would have a better performance than ESSnuSB.
\section{Conclusion}
The European Spallation Source Linac will be ready by 2023. ESS will have enough protons to go to the second oscillation maximum where the CPV sensitivity is significantly better. The preferred range of distances from the neutrino source to the detector site for CP violation discovery, is found to be between 300 km and 550 km. CPV: 5 $\sigma$  could be reached over 60  $\%$  of  $\Delta$$\delta$\textsubscript{CP} range (ESSnuSB) with large potentiality. COST network project CA15139 supports this project. ESSnuSB Design Study is approved by European Commission as a H2020 project and started on 1 January 2018.

\end{document}